\documentclass{article}%
\usepackage{amsmath}%
\usepackage{amsfonts}%
\usepackage{amssymb}%
\usepackage{graphicx}
\usepackage[dvips,final]{epsfig}
\usepackage[latin1]{inputenc}

\makeatletter

\newenvironment{figurehere}
  {\def\@captype{figure}}
  {}
\makeatother

\begin{document}

\title{Cosmological Equations for Interacting Energies}
\author{Alberto C. Balfagón(1), Raúl Ramírez-Satorras(2) 
\\and Ángel R. Martínez(3).
\\
\\Sección de Física Teórica y Aplicada. 
\\Departamento de Ingeniería Industrial,
\\Instituto Químico de Sarriá, Barcelona (Spain)
\\Vía Augusta 390, Barcelona, Spain
\\
\\(1)	albert.balfagon@iqs.edu
\\(2) raulramirezs@iqs.edu
\\(3) angelrodriguezm@iqs.edu
}
\date{}
\maketitle

\begin{abstract}

In this paper the coupling between dark energy and the other components of the cosmological fluid has been studied. Firstly, it will be shown that the application of general cosmological equations, deduced by the authors in a previous work, to the known data of the Abell cluster A586 using the Layzer-Irvine theory gives similar results compared to the work of other authors. The aforesaid method present some problems: the application of an approximate theory (Layzer-Irvine theory), it has only one experimental datum and finally, it gives results that are a bit difficult to admit considering a physical reasoning.
In order to avoid the above-mentioned problems, a way to study the coupling of dark energy with other Universe components has been shown. The results obtained have a sensible physical behavior. They also fix the required functionality of the product $w_\Lambda \Omega_\Lambda$ in order to verify the main known properties of the Universe's behavior. Finally, these results permit to make predictions about a set of different cosmological properties.
\\
\begin{flushleft}
\end{flushleft}
\end{abstract}

\tableofcontents

\section{Introduction}

Some experimental data obtained at the end of the last century, have showed that our Universe is speeding up its expansion. Recently, it has been found out that more accurate experimental data reach the same conclusion [1-4]

Several solutions, such as the existence of cosmological constant [5,6], modified theories of Gravity [7], dynamical dark energy [8-10] and unified dark energy-dark matter models [11], in order to explain the experimental data were tried . But, generally speaking, all these models had to face several problems, being the main one the coincidence problem and the fine tune of model parameters.

The newest models have started to study new possibilities like interactions between different kinds of components of the Universe, mainly interactions between dark energy and dark matter [12-20], new equations of state and mass-varying cold dark matter [21].

A more general model was developed recently by A.C. Balfagón and R. Ramírez-Satorras [22]. In this model some of the assumptions taken into account by  most authors, were relaxed. For instance, not considering that all the Universe's components had no interaction pressure between them. Within this recent model, the authors also studied an effective equation of state for the whole Universe instead of taking individual equations of state for each of its components.

This paper is focused on the study of the interactions between dark energy and the other Universe components, a part from the dark matter. Most of the models developed until the present time, have studied the dark energy and dark matter coupling; some authors have concluded that there is a energy transfer from dark matter to dark energy, while others stand for the opposite. Thus this is an issue involving much controversy [23-28].

One of the main problems of the interacting models is that, until this moment, the interaction has to be fixed beforehand, being the so called scaling interaction one of the most popular [29-31]. In this paper a scaling coupling and also a new form of the interaction that comes directly from our model have been studied. The latter taken into account a sensible physical assumption which has never been described before.

The rest of the paper is divided in six sections; section two exposes the general equations [22], and deduction of new ones. The third section deals with a scaling solution, similar to the one used for many authors but using a more general reading. The Layzer-Irvine theory is applied in order to find the parameters of the coupling using the experimental data from the Abell cluster A586. 

In the fourth section, a sensible physical condition is applied to the general equations deduced in section two; from this a new coupling function of dark energy and dark matter is deduced. In the section fifth different examples, deduced from the model of section fourth, have been studied. Finally, in the last section, the main conclusions are exposed.

\section{General equations.}

In [22] general equations for the Universe were deduced. The main idea behind the model was to take into account that the components of the Universe could interact between them. Thus it was important to consider some kind of interaction pressure between these components. With this idea, the application of the deduced equations can be done through the main history of the Universe. The main equations are:

\begin{equation}
	\rho'=\frac{3c^2}{8\pi G}\left(\frac{\dot{a}}{a}\right)^2-\frac{c^4}{8\pi G}\Lambda
\end{equation}

Where $\rho'$ stands for any kind of energy but the one that come from $\Lambda$, which can be considered as a constant (the so-called cosmological constant) or as a dynamical function.

Dark energy density $\rho_\Lambda$ will be:

\begin{equation}
	\rho_\Lambda=-\frac{c^4}{8\pi G}\Lambda
\end{equation}

The total energy density $\rho$ of the Universe, including dark energy, will be defined as:

\begin{equation}
	\rho=\rho'+\frac{c^4}{8\pi G}\Lambda=\rho'+\rho_\Lambda=\frac{3c^2}{8\pi G}\left(\frac{\dot{a}}{a}\right)^2
\end{equation}

The total pressure $p$ of the Universe, without assuming that this pressure is only the sum of the partial pressure components, will be:

\begin{equation}
	p=-\frac{c^2}{8\pi G}\left(2\frac{\ddot{a}}{a}+\left(\frac{\dot{a}}{a}\right)^2\right)
\end{equation}

Equation (5) is the effective equation of state (EOS) for the cosmological fluid (sum of all components). Equation (6) is the function for $w$ proposed in [22]. 

\begin{equation}
	p=w\rho
\end{equation}

\begin{equation}
	w=\frac{1}{3}\frac{\left(1+z\right)}{\left(\alpha+z\right)}-\gamma\frac{\left(\beta-1\right)}{\left(\beta+z\right)}
\end{equation}
  
To get the best set of parameters in Eq.(6),  a full factorial design of experiments (DOE) [22] was implemented using the data reported in [1].

\begin{equation}
	\alpha= 4.1, \beta= 3.5, \gamma= 1.0, h= 0.697, H(0)=100h
\end{equation}

With the above equations, the following equations can be easily deduced:

\begin{equation}
	w=\frac{1}{3}\frac{\left(1+z\right)}{\left(\alpha+z\right)}-\frac{\left(\beta-1\right)}{\left(\beta+z\right)}
\end{equation}

\begin{equation}
	\rho=\rho_0\left(\frac{z+\beta}{\beta}\right)^3\frac{\left(\alpha+z\right)}{\alpha}
\end{equation}

Where $\rho_0$ is the present fluid cosmological energy density.

\begin{equation}
	H=H_0\left(\frac{z+\beta}{\beta}\right)^{3/2}\left(\frac{\alpha+z}{\alpha}\right)^{1/2}
\end{equation}

The dark energy density variation can be deduced from Eq.(1), expressed now using the Hubble function $H$ as:

\begin{equation}
	H^2=\frac{8\pi G}{3c^2}\rho'+\frac{c^2}{3}\Lambda
\end{equation}

The most general variation of $\Lambda$ will be, using Eq.(11):

\begin{equation}
	\Lambda=\Phi H^2 +\Psi \rho'+\Lambda_0
\end{equation}

Where $\Phi,\Psi$ and $\Lambda_0$ are considered, in general, not constant. It is easy to show that the general expression (12) is equivalent to:

\begin{equation}
	\Lambda=\alpha_1\rho'+\alpha_0
\end{equation}

Consequently the dark energy density can be expressed as:

\begin{equation}
	\rho_\Lambda=\frac{c^4}{8\pi G}\Lambda=\beta_1\rho'+\beta_0
\end{equation}

In the last equation $\beta_1$ and $\beta_0$ are considered as some kind of effective coupling between dark energy and the others Universe components. The fact that this model includes any kind of coupling apart the one between dark energy and dark matter, makes it a newer and more global model.

With the present data, $\beta_0$ and $\beta_1$, have currently the values $0$ and $18/7$.

In the next sections some important models will be studied using these general equations.

\section{Scaling solution.}

In this section, a scaling solution for $\beta_1$ in Eq.(14) is studied. This kind of coupling has been previously studied by other authors [29-31]; however, considering it only as a coupling between dark energy and dark matter. In the model described in this article, this scaling solution is taken as an effective coupling between dark energy and the other components of the cosmological fluid.
The $\beta_1$ function will be:

\begin{equation}
	\beta_1=\frac{\delta}{(1+z)^\eta}
\end{equation}

In order to compare our results with the ones in [32] we will consider the present value of the dark energy density parameter $\Omega_\Lambda$ equal to $0.72$, using this the value for $\delta$ has to be $18/7$ and $\beta_0$ has to be $0$.
Substituting Eq.(14) in Eq.(3) the following is obtained:

\begin{equation}
	\rho_\Lambda=\frac{\beta_1}{1+\beta_1} \rho
\end{equation}

And substituting Eq.(15) in Eq.(16) the dark energy density evolution is deduced:

\begin{equation}
	\rho_\Lambda=\frac{18}{18+7(1+z)^\eta} \rho
\end{equation}

In order to get the value of the constant $\eta$ it is considered that in the present period:

\begin{equation}
	\rho\approx\rho_\Lambda+\rho_{dm}
\end{equation}

Where $\rho_{dm}$ stands for dark matter density

With Eq.(17) replaced in Eq.(18) the dark matter density is deduced to have the form:

\begin{equation}
	\rho_{dm}=\frac{7(1+z)^\eta}{18+7(1+z)^\eta} \rho
\end{equation}

Replacing Eq.(9) in Eq.(19) we get the dark matter density evolution as a function of the shift parameter $z$:

\begin{equation}
	\rho_{dm}=\frac{7\rho_0}{\alpha \beta^3} \frac{(z+\beta)^3 (\alpha+z) (1+z)^\eta}{18+7(1+z)^\eta}
\end{equation}

Following the procedure developed in [32], and applying the Layzer-Irvine equation [33] to the Abell Cluster A586, a value for $\eta$ can be obtained. 

\subsection{Layzer-Irvine equation}

The Layzer-Irvine equation is the equation for the classical virial equilibrium, applied for non-relativistic self-gravitating dust-like particles. It will be assumed that the mass of the system remains constant over the observation process.
All the matter of the system is assumed to be mainly dark matter. The local energy density for the dark matter will be the sum of the kinetic energy density and the self-gravitating energy density:

\begin{equation}
	\rho_{dm}^{'} =\rho_k +\rho_{dm}
\end{equation}

The $\rho_k$ can be computed considering the Newtonian kinetic energy $K$ per unit mass as:

\begin{equation}
	MK=\frac{1}{2a^2} \left\langle \frac{p^2}{m}\right\rangle
\end{equation}

Where the averaged momentum $p$ and mass $M$ are constants, $a$ is the scale factor of the Robertson-Walker's metric.
The $\rho_k$ will be:

\begin{equation}
	\rho_k=M\frac{dK}{dV}
\end{equation}

The $\rho_W$ is evaluated through the potential energy per unit mass:

\begin{equation}
	W=-2\pi G a^3 \rho_{dm} \int r\xi(r) dr
\end{equation}

Where $\rho_{dm}$ is taken as the background dark matter density, given by Eq.(20), and $\xi$ is the auto-correlation function.
The $\rho_W$ will be:

\begin{equation}
	\rho_W=M\frac{dW}{V}
\end{equation}

The derivation of Eq.(21) gives the Layzer-Irvine equation:

\begin{equation}
	\frac{d \rho_{dm}^{'}}{dt}=-\left[2\rho_K+(1+z)\zeta \rho_W\right] H
\end{equation}

Where $H$ is the Hubble function, and $\zeta$ is given by:

\begin{equation}
	\zeta=\frac{3}{z+\beta}+\frac{1}{z+ \alpha}+\frac{\eta -2}{1+z}-\frac{7 \eta (1+z)^{\eta-1}}{18+7(1+z)^{\eta}}
\end{equation}

If the system is stationary then $\dot{\rho}_{dm}^{'}=0$, then from Eq.(26):

\begin{equation}
	2\rho_K=-(1+z)\zeta \rho_W
\end{equation}
 
\subsection{The Abell cluster A586}

The Abell cluster A586 can be considered as an stationary system, where Eq.(28) can be applied, with a redshift of $z=0.1708$. The kinetic and gravitational energy densities can be estimated as in [32]. The quotient value and band error are:

\begin{equation}
	\frac{\rho_K}{\rho_W}\approx-0.76\pm0.05
\end{equation}

Using these data in Eq.(27) and Eq.(28) it is possible to get a value for $\eta$, this value is $3.95\pm0.24$, while in [32] the value was $3.82$. These values indicate that the coupling, at least in the recent history of the Universe, between dark energy and dark matter is inducing a transfer of energy from dark matter to dark energy [32].

In Fig.1, the dark energy density evolution for $z\in[-1,4]$ using Eq.(17) and Eq.(9) is shown.

\begin{figurehere}
	\centering
	\vspace{0.5cm}
	\hspace{0.5cm}
	\includegraphics[width=10cm]{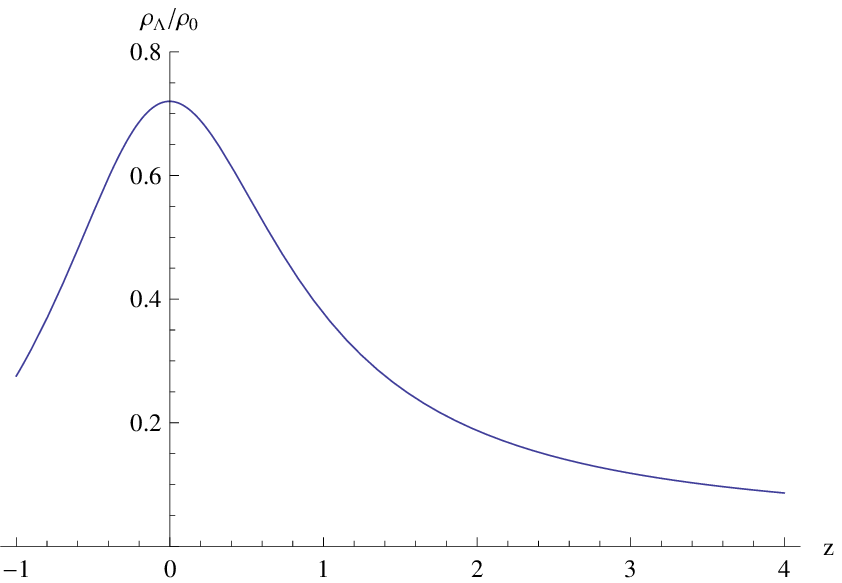}
	\caption{ $\frac{\rho_\Lambda}{\rho_0}$ against $z$}
	\vspace{0.5cm}
	\label{fig:fig1}
\end{figurehere}

It seems clear that the dark energy density behaves in a bit strange way, why it is maximum for $z\approx0$?. In the next section the coupling function given by Eq.(15) will be changed in order to solve this problem.

In Fig.2, the dark matter density evolution for $z\in[-1,4]$ from Eq.(20) is shown.

\begin{figurehere}
	\centering
	\vspace{0.5cm}
	\hspace{0.5cm}
	\includegraphics[width=10cm]{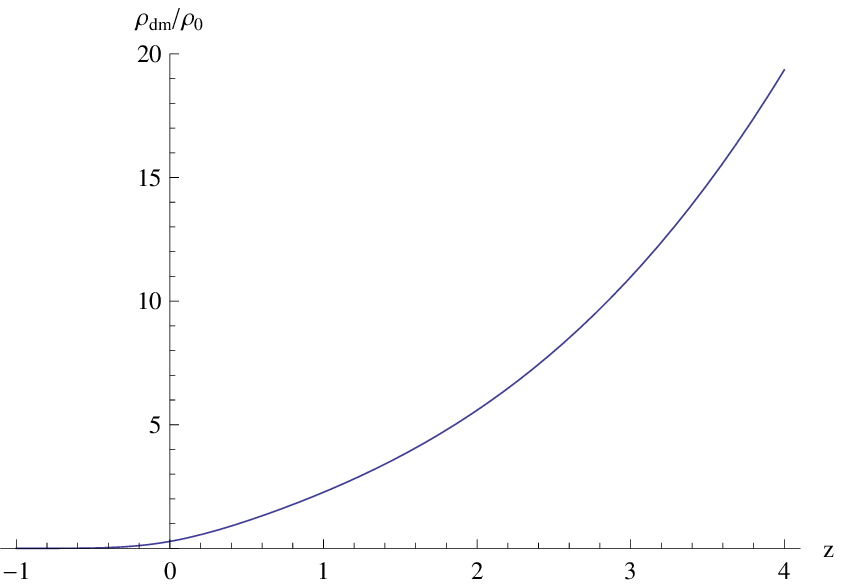}
	\caption{ $\frac{\rho_{dm}}{\rho_0}$ against $z$}
	\vspace{0.5cm}
	\label{fig:fig1}
\end{figurehere}

Furthemore using Eq.(4) it is possible to see if this dark sector coupling it is enough to explain all possible recent interactions between the components of the Universe. The pressure in Eq.(4) can be considered as a sum of several partial pressure added to a generic interaction pressure:

\begin{equation}
	p=w\rho=p_m+p_r+p_\Lambda+p_i
\end{equation}

Where $p_m$ is the matter pressure, $p_r$ the radiation pressure, $p_\Lambda$ the dark energy pressure , $p_i$ the interaction pressure and $w$ is given by Eq.(6). Considering the equation of state usually taken for the above components we have:

	\[p_m=0
\]

\begin{equation}
	p_r=-\frac{0.0001}{3}(1+z)^4
\end{equation}
	
	\[p_\Lambda=-\rho_\Lambda
\]

Where $\rho_\Lambda$ is given by Eq.(17).

The interaction pressure is deduced replacing the above equations in Eq.(30):

	\[
	p_i=\rho_0\left(\frac{z+\beta}{\beta}\right)^3\frac{(\alpha+z)}{\alpha}\left(\frac{1+z}{3(\alpha+z)}-\frac{\beta-1}{\beta+z}+\frac{18}{18+7\left(1+z\right)^\eta}\right)
\]
\begin{equation}
-\rho_0\frac{0.0001}{3}\left(1+z\right)^4	
\end{equation}

In Fig.3 the Eq.32 for $z\in[-1,4]$ is shown.

\begin{figurehere}
	\centering
	\vspace{0.5cm}
	\hspace{0.5cm}
	\includegraphics[width=10cm]{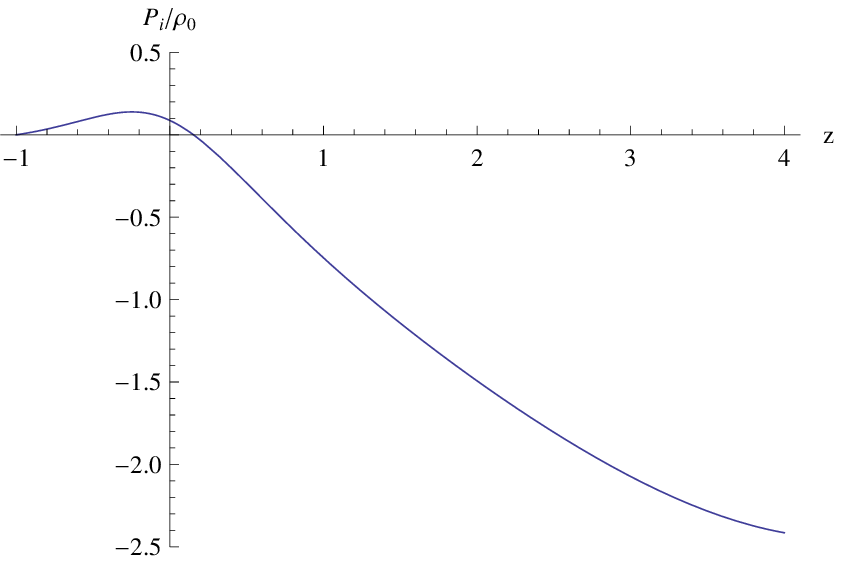}
	\caption{ $\frac{p_i}{\rho_0}$ against $z$}
	\vspace{0.5cm}
	\label{fig:fig1}
\end{figurehere}

Watching Fig.3, it seems to be clear that, in the recent history of the Universe (for $z$ less or equal to 3) there is still some room for other interactions, or that the coupling effective function given by Eq.(15) doesn´t work properly in order to explain the coupling in the dark sector. An alternative way to explain all the interactions for the recent history is showed in the next section.

\section{General coupling function.}

In this section, the possibility of finding a coupling function between dark energy and the other components of the Universe using a sensible physical condition, will be described.

The procedure consists in considering Eq.(16) but without specifying any kind of function for $\beta_1$. Then Eq.(16) is expressed as:

\begin{equation}
	\rho_\Lambda=\frac{\beta_1}{1+\beta_1} \rho = A \rho
\end{equation}

Where, obviously, $A$ is the adimensional density parameter $\Omega_\Lambda$.

The EOS for the cosmological fluid is given by Eq.(30), where $w$ is given by the Eq.(8), $\rho$ is the Eq.(9) and $p_m$ and $p_r$ are given by Eq.(31). The EOS for the dark energy is now taken as:

\begin{equation}
	p_\Lambda=w_\Lambda \rho_\Lambda = w_\Lambda A \rho
\end{equation}

Replacing the above equations in Eq.(30) the interaction pressure is deduced as follows:

	\[
	p_i=\rho_0 \frac{(z+\beta)^3}{\beta^3}\frac{(\alpha+z)}{\alpha}\left(\frac{1+z}{3(\alpha+z)}-\frac{\beta-1}{\beta+z}-w_\Lambda A\right)
\]
\begin{equation}
	-\rho_0 \frac{0.0001}{3}\left(1+z\right)^4
\end{equation}

At this point the physical assumption of considering that, for $z\leq 3$, the interaction pressure between the components of the Universe is meaningless gives rise to:

\begin{equation}
	w_\Lambda A= \frac{1+z}{3(\alpha+z)}-\frac{\beta-1}{\beta+z}-\frac{\alpha \beta^3 0.0001(1+z)^4}{3(z+\beta)^3 (\alpha+z)}
\end{equation}

It is worthwhile noticing that for $z = 0$ Eq.(36) is equal to $-0.63$, while the most probable value for the adimensional density parameter is $0.72$ with $w<0$.

In Fig.4 the evolution of Eq.(36) for $z\leq 3$ is shown.

\begin{figurehere}
	\centering
	\vspace{0.5cm}
	\hspace{0.5cm}
	\includegraphics[width=10cm]{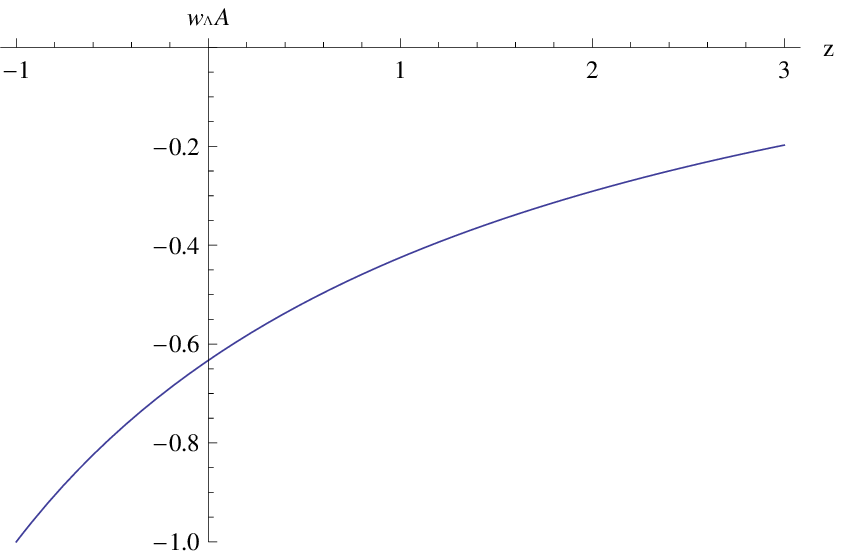}
	\caption{ $w_\Lambda A$ against $z$}
	\vspace{0.5cm}
	\label{fig:fig1}
\end{figurehere}

The high degeneracy between $w_\Lambda$ and $\rho_\Lambda$ is remarkable. In the next subsections two important cases will be discussed.

\subsection{$w_\Lambda=-1$}

If $w_\Lambda$ is assumed to be constant, at least for $z\leq3$ and equal to $-1$, then from Eq.(36) the adimensional density parameter $\Omega_\Lambda(z)$ can be deduced:

\begin{equation}
	\Omega_\Lambda(z)=A= -\frac{1+z}{3(\alpha+z)}+\frac{\beta-1}{\beta+z}+ \frac{\alpha \beta^3 0.0001(1+z)^4}{3(z+\beta)^3 (\alpha+z)}
\end{equation}

From Eq.(37) it can be obtained that $\Omega_\Lambda(0)=0.63$.

At this moment, the $\rho_\Lambda$ is given from Eq.(33) and Eq.(37):

\begin{equation}
	\rho_\Lambda=A \rho_0 \frac{(z+\beta)^3}{\alpha \beta^3}(\alpha+z)
\end{equation}

Fig.5 shows the evolution of $\rho_\Lambda$. Comparing this evolution with Fig.1 it is evident that the behavior of Eq.(38) is more physically sensible.

\begin{figurehere}
	\centering
	\vspace{0.5cm}
	\hspace{0.5cm}
	\includegraphics[width=10cm]{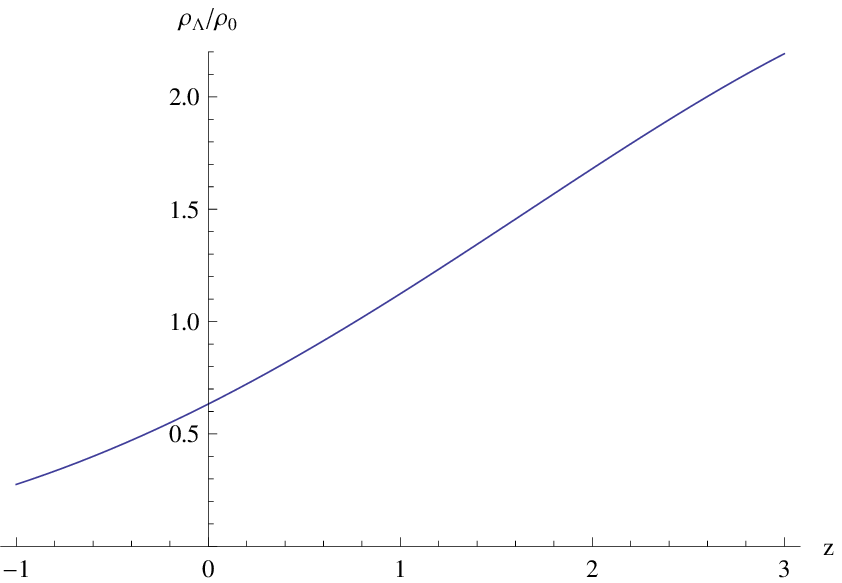}
	\caption{ $\rho_\Lambda/\rho_0$ against $z$}
	\vspace{0.5cm}
	\label{fig:fig1}
\end{figurehere}

From the equations (33) and (37) $\beta_1$ is easily deduced. In Fig.6 the evolution of $\beta_1$ for $z\leq3$ is revealed.

\begin{figurehere}
	\centering
	\vspace{0.5cm}
	\hspace{0.5cm}
	\includegraphics[width=10cm]{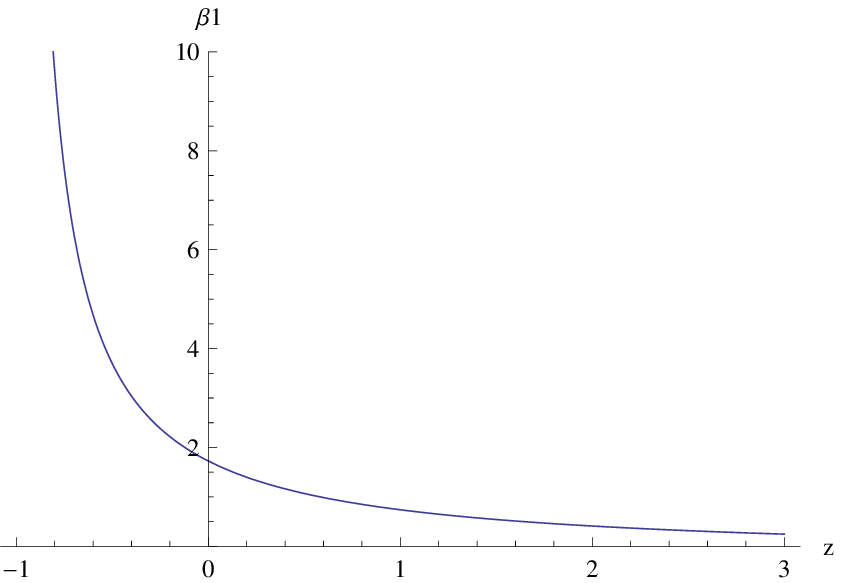}
	\caption{ $\beta_1$ against $z$}
	\vspace{0.5cm}
	\label{fig:fig1}
\end{figurehere}

\subsection{Variable scaling function.}

From Eq.(36), and taking $\Omega_\Lambda(0)=0.72$, the $w_\Lambda$ deduced is $-0.88$. Taking $w_\Lambda$ as a constant, at least for $z\leq3$, the equations for $\Omega_\Lambda$ and for $\rho_\Lambda$ are very similar to equations (37) and (38).

In this case it is possible to match Eq.(17) with Eq(33) taking for the $A$ function the one deduced from Eq.(36) with $w_\Lambda=-0.88$. Both equations give rise to the same value for $\Omega_\Lambda(0)$. For the match to be possible it is necessary to consider the variable $\eta$ not constant, given a variable scaling function:

\begin{equation}
	\eta=\frac{\ln(\frac{18}{7}\left(\frac{1}{A}-1\right))}{\ln(1+z)}
\end{equation}

Evaluating Eq(39) for the Abell cluster A586 redshift $z=0.1708$, a value can be deduce for $\eta$ equal to $1.48$; as it can be observed, this is a very different value from the one deduce in [32] and from equations (28) and (29). This value for $\eta$ indicates that there is a transfer from dark energy to the other components of the universe.

In Fig.7,the evolution of $\eta$, Eq.(39) for $z\in\left[-0.1, 3\right]$ is shown. It should be pointed out that the variable $\eta$ is almost constant and less than $3$ (the limit value that implies a transfer from dark energy to the other components of the Universe $\eta<3$ or vice versa $\eta>3$, thus the coupling between dark energy and the rest of cosmological fluid seem to be inducing a transfer of dark energy towards the other components, at least for $z\leq3$.

\begin{figurehere}
	\centering
	\vspace{0.5cm}
	\hspace{0.5cm}
	\includegraphics[width=10cm]{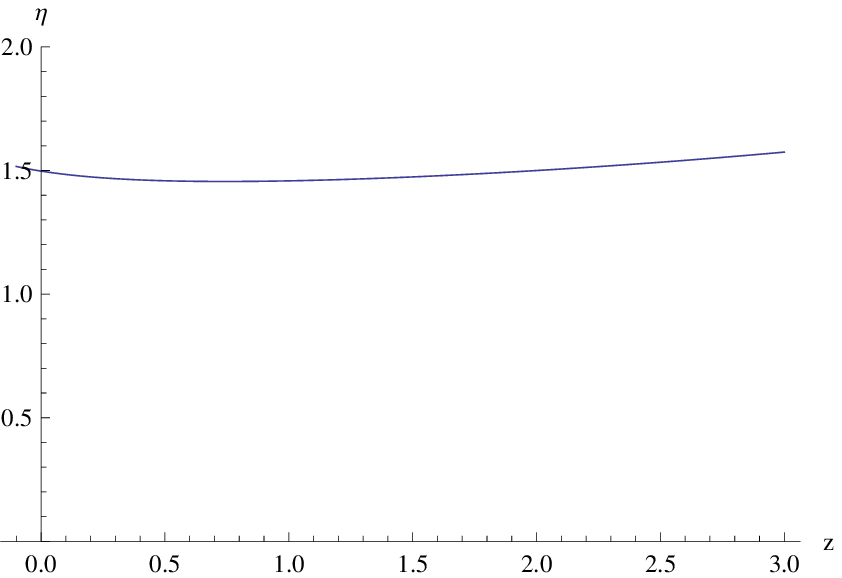}
	\caption{ $\eta$ against $z$}
	\vspace{0.5cm}
	\label{fig:fig1}
\end{figurehere}

In Fig.8 simply shows a magnification of Fig.7.

\begin{figurehere}
	\centering
	\vspace{0.5cm}
	\hspace{0.5cm}
	\includegraphics[width=10cm]{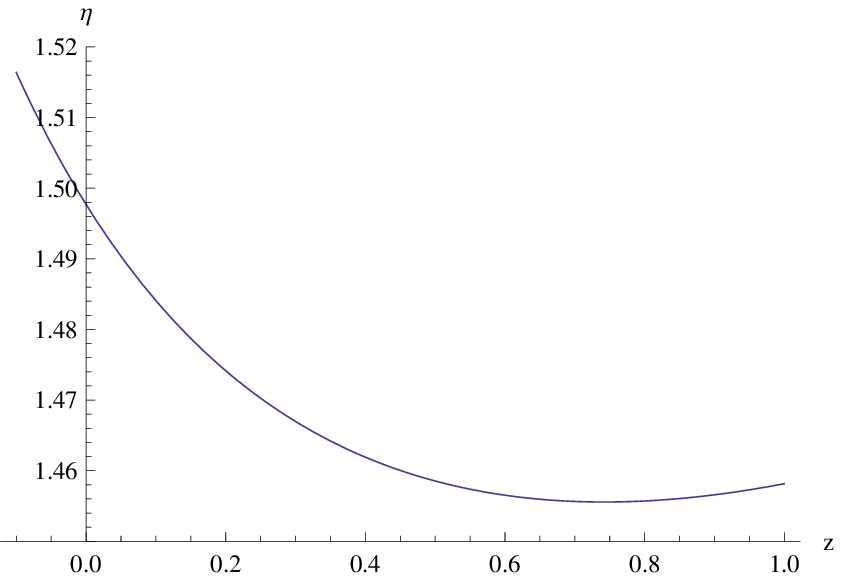}
	\caption{ $\eta$ against $z$}
	\vspace{0.5cm}
	\label{fig:fig1}
\end{figurehere}

\section{Conclusions.}

In this paper, the application of a general cosmological model developed by the authors previously [22] has been taken into account to study the coupling between dark energy and the other components of the Universe.

Initially, an arbitrary coupling function of the scaling type was assumed. The method used, in order to find out the value of the scaling constant $\eta$, was the one developed in [32] using the Abell cluster A586 data and the Layzer-Irvine theory. The results obtained were very similar to the ones deduced in [32]. The value for $\eta$ was $3.95$ which implies that the coupling is transferring energy to the dark energy from the other Universe components. Using such coupling function, it is clear that there is still some room for other couplings, and that the evolution of the dark energy density has a behavior a bit unusual: it grows up to a maximum near $z=0$ and then goes down. Generally speaking this method has a lack of experimental data and suffers from the theory applied: the arbitrary coupling function used and the classical Layzer-Irvine theory applied.

In order to avoid the problems of the above-mentioned method, in the second part of the paper the application of a sensible physical condition was considered. The assumption adopted was considering that for $z\leq3$ the interaction pressure between the cosmological components was negligible and from this it was deduced the necessary coupling function in order to fulfill this condition.

The function deduced from the above condition was the product of $w_\Lambda$ and $\Omega_\Lambda$, which shows the high degeneracy for $w_\Lambda$ and for $\Omega_\Lambda$. This function has some relevant features that made it interesting for the research in dark energy:

\begin{itemize}
	\item 
The value of the function for $z=0$ is $w_\Lambda \Omega_\Lambda = -0.63$, while the best experimental value for $\Omega_\Lambda$ is $0.72$.
  \item
It gives the general form of the product $w_\Lambda \Omega_\Lambda$ in order to fulfill two main conditions: the accelerated expansion of the Universe and the negligible value of the present interaction pressure.	
\end{itemize}

If $w_\Lambda$ is assumed to be constant, at least for $z<3$, then two relevant cases are easily studied:

\begin{itemize}
	\item 
Taking $w_\Lambda=-1$, it is predicted that $\Omega_\Lambda(0)=0.63$. It is also deduced a new coupling function, $\beta_1$, between dark energy and the other components of the Universe, and the $\Omega_\Lambda$ function $(\Omega_\Lambda=A)$. From these two functions a more sensible function for the dark energy density is deduced.  
  \item
Taking $\Omega_\Lambda(0)=0.72$, the value predicted for $w_\Lambda$ is $-0.88$. In this case the $\beta_1$, $\Omega_\Lambda$ and dark energy density functions are similar to the above case with $w_\Lambda=-1$. However, now it is possible to deduce a new scaling coupling function with $\eta$ variable. In this case it is shown that the coupling is transferring dark energy to the other components of the Universe, at least for the z interval where the assumption that was made is applicable. Furthermore, $\eta$ varies very little for $z\in\left[0,3\right]$.  
\end{itemize}

\end{document}